\documentclass[twocolumn,twoside,slac_two]{revtex4}
\usepackage{graphicx}
\usepackage{fancyhdr}
\usepackage{sidecap}
\usepackage{gensymb} \usepackage{subfigure}
\DeclareRobustCommand{\orderof}{\ensuremath{\mathcal{O}}}
\RequirePackage{lineno}
\usepackage{color,soul}
\usepackage{amsmath}

\newcommand{\Fermi}{{\it Fermi }}

\skip\footins=1.5\bigskipamount     
\renewcommand*{\footnoterule}{%
  \kern-3pt%
  \hrule width 2in%
  \kern 2.6pt%
  \vspace{\smallskipamount}       
}

\def\apj{ApJ }				
\def\aap{A\&A } 				
\def\bams{\ref@jnl{Bull.~Am.~Meteorol.~Soc.}} 
\def\bssa{\ref@jnl{Bull.~Seismol.~Soc.~Am.}} 
\def\dsr{\ref@jnl{Deep~Sea~Res.}}	
\def\eos{\ref@jnl{Eos~Trans.~AGU}}	
\def\epsl{\ref@jnl{Earth~Planet.~Sci.~Lett.}}	
\def\gca{\ref@jnl{Geochim.~Cosmochim.~Acta}}	
\def\gjras{\ref@jnl{Geophys.~J.~R.~Astron.~Soc.}} 
\def\grl{\ref@jnl{Geophys.~Res.~Lett.}}	
\def\gsab{\ref@jnl{Geol.~Soc.~Am.~Bull.}}	
\def\jatp{\ref@jnl{J.~Atmos.~Terr.~Phys.}}	
\def\jgr{\ref@jnl{J.~Geophys.~Res.}}	
\def\jpo{\ref@jnl{J.~Phys.~Oceanogr.}}	
\def\mnras{MNRAS}			
\def\mwr{\ref@jnl{Mon.~Weather~Rev.}}	
\def\nat{Nature}
\def\pepi{\ref@jnl{Phys.~Earth Planet.~Inter.}}	
\def\pra{\ref@jnl{Phys.~Rev.~A}}		
\def\prb{\ref@jnl{Phys.~Rev.~B}}		
\def\prc{\ref@jnl{Phys.~Rev.~C}}		
\def\prd{\ref@jnl{Phys.~Rev.~D}}		
\def\prl{\ref@jnl{Phys.~Rev.~Lett}}	
\def\qjrms{\ref@jnl{Q.~J.~R.~Meteorol.~Soc.}}	
\def\rg{\ref@jnl{Rev.~Geophys.}}	
\def\rs{\ref@jnl{Radio~Sci.}}		
\def\usgsof{\ref@jnl{U.S.~Geol.~Surv. Open File~Rep.}}	
\def\usgspp{\ref@jnl{U.S.~Geol.~Surv.~Prof.~Pap.}}	

\begin{document}

\title{Observations of the Crab Pulsar with VERITAS}

%

\author{A. McCann\footnote{mccann@kicp.uchicago.edu} for the VERITAS Collaboration}
\affiliation{The University of Chicago, The Kavli Institute for Cosmological Physics, 933 East 56th Street, Chicago, IL 60637}

\begin{abstract}
The \emph{Fermi} space telescope has detected over 100 pulsars. These
discoveries have ushered in a new era of pulsar astrophysics at
gamma-ray energies. Gamma-ray pulsars, regardless of whether they are
young, old, radio-quiet etc, all exhibit a seemingly unifying
characteristic: a spectral energy distribution which takes the form of
a power law with an exponential cut-off occurring between $\sim$1 and
$\sim$10 GeV. The single known exception to this is the Crab pulsar,
which was recently discovered to emit pulsed gamma rays at energies
exceeding a few hundred GeV. Here we present an update on observations
of the Crab pulsar above 100 GeV with VERITAS. We show some new
results from a joint gamma-ray/radio observational campaign to search
for a correlation between giant radio pulses and pulsed VHE emission
from the Crab pulsar. We also present some preliminary results on
Lorentz invariance violation tests performed using \emph{Fermi} and
VERITAS observations of the Crab pulsar.
\end{abstract}

\maketitle

\thispagestyle{fancy}


\section{Introduction}
The \Fermi space telescope has detected over 100 new gamma-ray
pulsars, the bulk of which are young, high-$\dot{E}$, pulsars. These
discoveries have ushered in a new era of research into pulsars and the
physics of emission from pulsar magnetospheres. However, despite the
wealth of new data in recent years, the decades-old problem of the
origin of gamma-ray emission from pulsars remains unsolved. The origin
of the coherent radio emission from pulsars is also poorly understood,
as are the exotic temporal phenomena of \emph{Nulling} and \emph{Giant Radio Pulses} (GRPs) exhibited by some radio pulsars.

In spite of the difficulty faced to explain the emission from pulsars,
their regular and predictable pulsations and immense density makes
them incredibly useful laboratories for the study of space-time and
gravitation. In this contribution, we present an update on
observations of the Crab Pulsar above 100~GeV with the Very Energetic
Ray Imaging Telescopes Array System (VERITAS). In particular, we focus
on a recently completed study to search for a correlation between the
emission of GRPs and pulsed very-high-energy (VHE) gamma-ray emission
and preliminary results from a study searching for the effects of
quantum gravity on the propagation of photons of different energies
over large distances.

\section{Searching for a VHE-GRP Correlation}
\begin{figure}
\centering \includegraphics[width=0.47\textwidth]{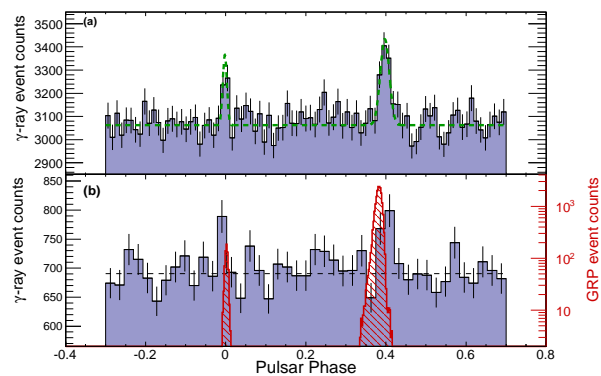}
\caption{~Panel (a) shows the VHE gamma-ray pulse profile ($E_{\gamma}
  > $ 120 GeV) of Crab pulsar measured by VERITAS (see
  \cite{Aliu2011Sci}). The overlaid (green) curve was determined from
  a maximum likelihood fit to the unbinned VERITAS data. Panel (b)
  shows the VERITAS (blue) and GRP (red) profiles for 11.6 hours
  of simultaneous VERITAS/GBT observations.  }
\end{figure}
\subsection{GRPs in the Crab Pulsar}
\begin{figure*}
\subfigure[$\,\,\,$ Event Arrival Times]
{
\includegraphics[height=5.2cm]{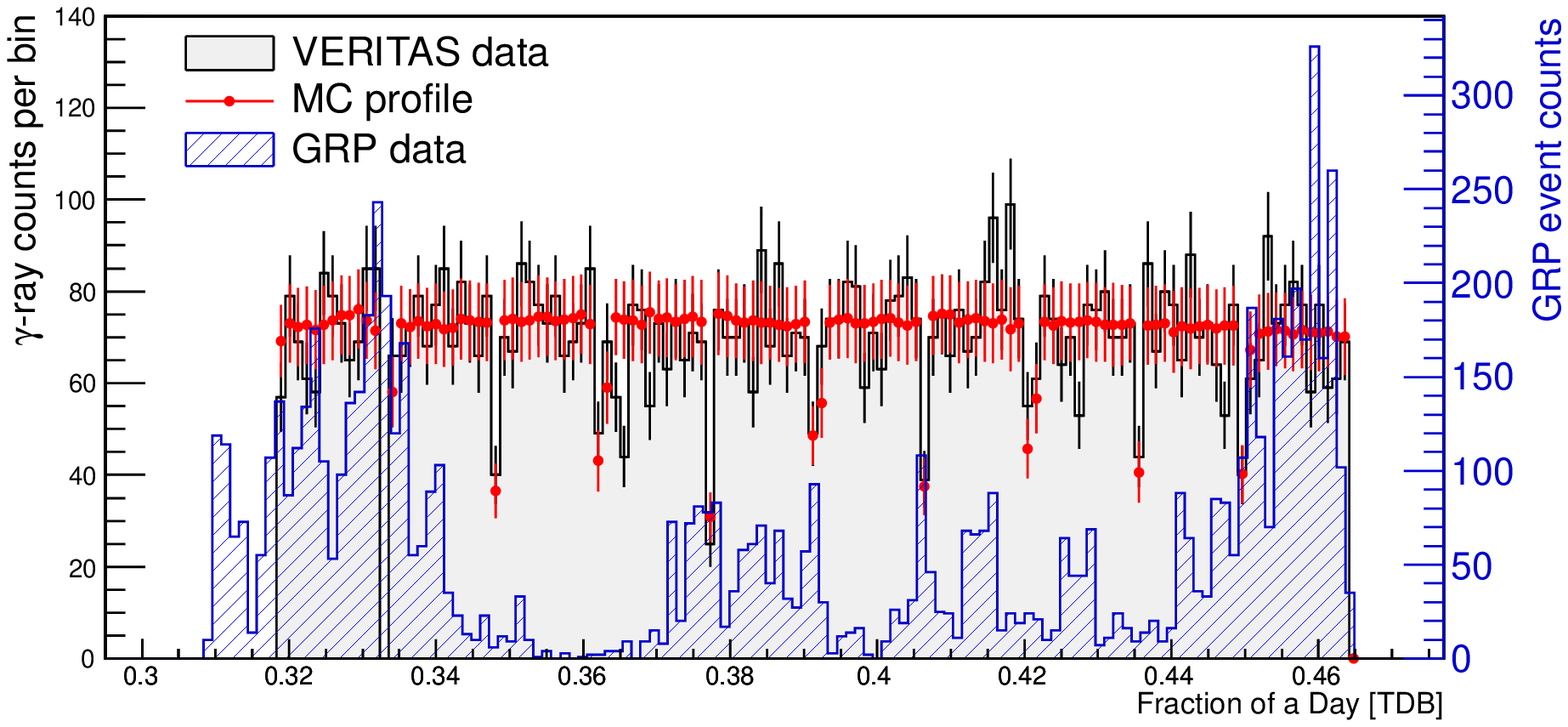}
}
\subfigure[$\,\,\,$Event Separation]
{
\includegraphics[height=4.75cm]{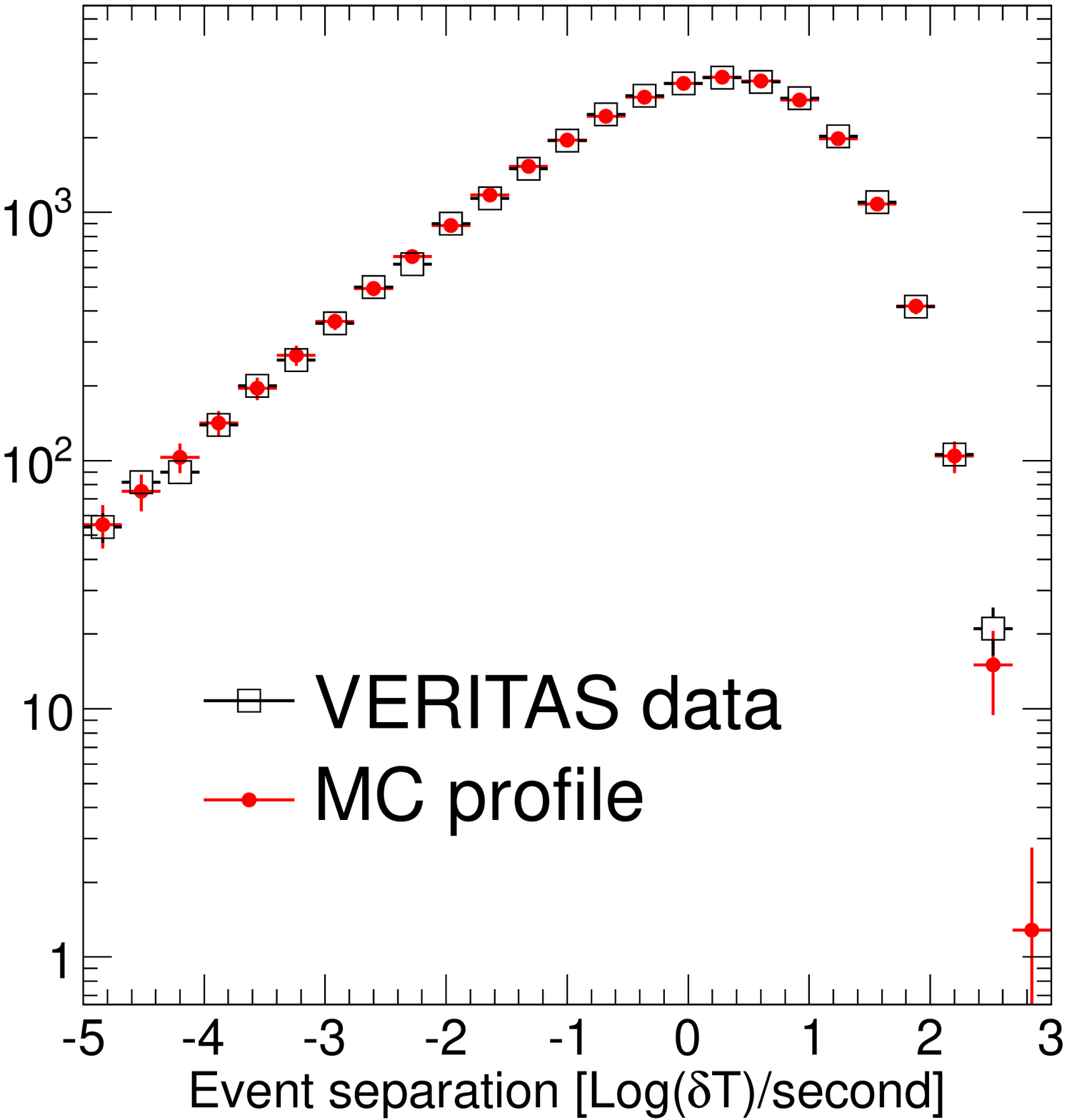}
}
\caption{This figure shows examples of the match between the VERITAS
  data and the Monte-Carlo model. The event rate distribution is shown
  on the left and the arrival time event-separation distribution for
  each pair of consecutive events is shown on the right.}
\end{figure*}
The Crab pulsar, PSR B0531+21, is a powerful young pulsar and is, so
far, the only pulsar to be detected above 100~GeV
\citep{Aliu2011Sci,MAGIC2011}. The Crab pulsar is also one of only
several known pulsars exhibiting the phenomenon of giant radio pulse
\citep{Knight06}: single radio pulses with flux densities that greatly
exceed the average pulse flux density. GRPs appear in the same phase
range as normal radio pulses, but their energy distribution is a
power-law \citep{Cordes2004,Popov2007} in contrast to the regular
pulses that follow a Gaussian or log-normal distribution
\citep{BurkeSpolaor2012}. In the Crab pulsar, individual GRPs can be a
few nanoseconds to a few microseconds wide \citep{Hankins2003Natur}
and can, at their maximum, be as bright as a few million Janskys
\citep{Soglasnov07}.

Above 6~GHz, stark differences in the behaviour of main pulse GRPs and
interpulse GRPs have been observed in the Crab pulsar. Interpulse GRPs
are typically several microseconds long and populate a set of
regularly spaced frequency bands \citep{Hankins2007} in contrast to
main pulse GRPs which exhibit broadband spectra and appear as a
succession of narrow pulses ranging from unresolved widths below
0.4~ns to widths of a few microseconds. In the model of
\cite{Lyutikov2007}, the band structure seen in the Crab pulsar
interpulse GRPs at high frequencies is generated by particles in the
outer pulsar magnetosphere which have been energised by a magnetic
field reconnection and emit via anomalous cyclotron resonance. A
prediction of this model is the existence of a particle beam in the
outer magnetosphere with a large Lorentz factor, $\orderof(10^{8})$,
which should generate curvature photons with energies as high as tens
of GeV. The mechanisms responsible for the generation of GRPs are,
however, still largely unknown.  Changes in the coherence of the
plasma beam, which is believed to be responsible for the normal pulsed
radio emission, can in principle explain the generation of GRPs. Such
coherence changes will, however, have no effect on the incoherent
emission from pulsars (optical, x-ray, gamma-ray).  Mechanisms which
increase the rate of particle production within the magnetospheric
emission region, or which change the direction of the emission beam
can, in principle, affect the higher energy incoherent emission and
may create an enhancement in the gamma-ray emission.

Several previous studies have been performed to search for a possible
connection between GRPs in the Crab and higher-energy incoherent
emission. While only upper-limits have been report at x-ray
(1.4-4.5~keV) \citep{Bilous2012} and gamma-ray (50-220~keV and
0.1-5~GeV) \citep{Lundgren1995,Bilous2011} energies,
\cite{Shearer2003}, observed a significant 3\% enhancement in the
optical main pulse concurrent with the period of emission of GRPs
measured at 1380 MHz. No enhancement was seen in the interpulse. This
observation suggests a link between coherent radio emission and
incoherent optical emission in the Crab pulsar. Small changes in the
pair-creation rate leading to localised density increases within the
emission plasma could create the GRP event and provide a small
enhancement in the optical incoherent emission.

\subsection{Observations and Search Strategy}
VERITAS and the 100-m Robert C. Byrd Green Bank Telescope (GBT)
simultaneously observed the Crab pulsar for a total of 11.6 hours
across four nights in December 2008 and November and December 2009.  A
total of 15366 GRP radio events recorded at 8.9~GHz, whose peak
signal-to-noise ratio was at least seven times greater than the
average radio signal, were selected for the correlation study (see
Figure~1). The VERITAS observations were made at high elevation ($El >
70^{\circ}$) under the best possible sky conditions with the four
telescopes fully operational and with a GPS time-stamp event tagging
error $< 10~\mu$s. The VERITAS event reconstruction and cosmic-ray
rejection followed the identical procedure as described in
\citep{Aliu2011Sci}.  The GPS time of the candidate gamma-ray events
which passed analysis cuts was converted to barycentric dynamical time
and phase-folded using the Crab pulsar monthly timing
ephemeris\footnote{\url{http://www.jb.man.ac.uk/pulsar/crab.html}}
\citep{Lyne1993}.

Using the VERITAS and GRP photon lists a correlation analysis was
performed. VERITAS events were selected if their barycentric arrival
time was within a time window positioned around the GRP arrival
times. Gamma-ray enhancements which lagged, lead or were
contemporaneous with the GRP were investigated. Since the correlation
length is unknown (if a correlation exists at all), eight different
coincidence time windows were chosen lasting; 1, 3, 9, 27, 81, 243,
729 and 2187 pulsar rotations. Main pulse and interpulse GRPs were
also considered separately, thus 72 different correlation searches
were performed. A Monte-Carlo model of the VERITAS data was used to
determine the number of coincident VHE events expected in the absence
of a correlation. Monte-Carlo time series data sets were generated by
drawing random times from the raw VERITAS trigger rate histograms. A
random subset of these events was chosen to contain the temporal
signature of the pulsar with phase values draw from the functional
form which describes the shape of the pulse profile of the Crab pulsar
as measured by VERITAS (see the dashed curve in panel (a) in
Figure~1). This procedure was used to generate sets of Monte-Carlo
data which contain a signal from the Crab pulsar at a chosen flux
level and which model all the data rate characteristics of the real
VERITAS data. Examples of the match between the Monte-Carlo
time-series datasets and the VERITAS gamma-ray data are shown in
Figure 2.

\subsection{GRP-VHE Correlation Results}
\begin{figure}
\centering \includegraphics[width=0.47\textwidth]{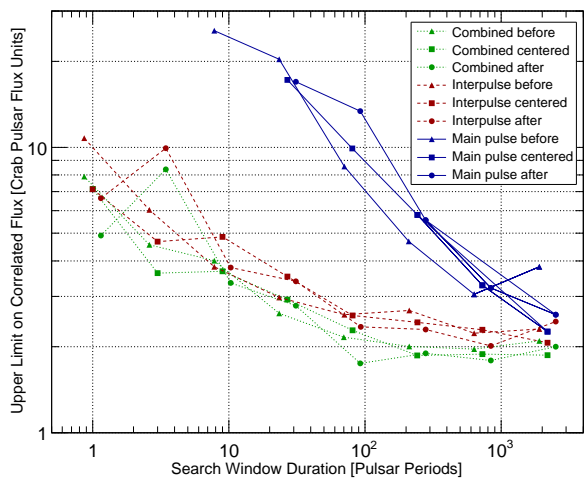}
\caption{~95\% confidence level upper limit on the VHE gamma-ray flux
  correlated with GRPs. The slight negative and positive shift of the
  x-position of the \textit{before} and \textit{after} symbols is done
  as a visual aid. Limits are given in units of the average VHE pulsar
  flux measured by VERITAS \citep{Aliu2011Sci}.}
\end{figure}
No significant enhancement in VHE gamma-ray emission correlated with
GRPs measured at 8.9~GHz was observed in any of the 72 searches
performed. Using the Monte-Carlo model of the Crab pulsar signal
within the VERITAS data, it is possible to investigate the probability
that a given number of correlated events is consistent with a chosen
level of correlated VHE enhancement. Following a similar procedure as
\cite{Bilous2011}, the 95\% confidence level upper-limit on the
enhanced VHE emission correlated with GRPs was calculated for each of
the 72 searches performed. These upper limit values are shown in
Figure~3.

These findings agree with results previously reported at keV and GeV
energies \citep{Bilous2012,Lundgren1995,Bilous2011}. GRP-emission
mechanisms associated with changes in plasma coherence will not cause
enhancements in incoherent emission, while small and localised changes
in the pair-creation rate can explain the small (3\%) optical-GRP
correlation seen by \cite{Shearer2003} but yield VHE enhancements
below the current VHE sensitivity. In recent models of pulsed VHE
emission from the Crab pulsar \citep{Aharonian2012,Lyutikov2011} any
expected GRP-VHE enhancements are also below the sensitivity of
current VHE instruments.

See \cite{Aliu2012} for a detailed description of this VHE-GRP
correlation study.

\section{Searching for Quantum Gravity}
Many theoretical investigations aim to formulate a theory of quantum
gravity in order to unify the theories of quantum mechanics and
general relativity. Some approaches to the theory of quantum gravity
result in equations of motion which contain an energy dependent
dispersion relation for massless particles, meaning that the speed of
light in a vacuum would not be constant. Such theories would violate
Lorentz invariance. Depending on the particular framework, photons
would travel a the speed
\begin{equation}
\nu(E)  = c\left[\,\,1 \pm \left(\frac{E}{E_{QG}}\right)^{n}\,\,\right]
\end{equation}
where $E$ is the photon energy and $E_{QG}$ is the energy scale at
which the effects of quantum gravity become important
\citep{Amelino1998}. The order, $n$, of the energy dependence, and the
$\pm$ sign ambiguity would be fixed within the framework of a given
model for quantum gravity. To search for these dispersive quantum
gravity effects one can apply the following framework:
\begin{itemize}
\item Two photons with energies $E_{1}$ and $E_{2}$ emitted
  simultaneously arrive at an observer at a distance, $L$, with time
  separation, $\Delta t$.
\item Depending on the order $n$ of the energy dependence, the time
  difference $\Delta t$ for the linear and the quadratic terms in
  ($E/E_{QG}$) are:
\vspace{-0.8cm}
\begin{center}
\begin{equation}
\Delta t_{1}  =  \frac{L}{c}\frac{E_{2}-E_{1}}{E_{QG}} \Rightarrow E_{QG} = \frac{L}{c}\frac{E_{2}-E_{1}}{\Delta t_{1}}
\end{equation}
\end{center}
\vspace{-0.5cm}
\begin{center}
and
\end{center}
\vspace{-1.3cm}
\begin{center}
\begin{equation}
\Delta t_{2}  =  \frac{L}{c}\frac{3}{2}\frac{E^{2}_{2}-E^{2}_{1}}{E^{2}_{QG}} \Rightarrow E_{QG} = \sqrt{\frac{L}{c}\frac{3}{2}\frac{E^{2}_{2}-E^{2}_{1}}{\Delta t_{2}}}
\end{equation}
\end{center}
\item By measuring the structure in a temporal event, such as an
  emission flare, in different energy bands, we can place an upper
  limit on $\Delta t$.
\item By placing a limit on the maximal $\Delta t$, a lower bound on
  $E_{QG}$ can be found.
\end{itemize}

From Equations 2 and 3 it is clear that when search- ing for Lorentz
invariance violation (LIV) it is favor- able to observe temporal
phenomena which occur on short time scales (minimising $\Delta t$) at a
large distance from the observer (maximising $L$) over a wide photon
energy range (maximising $E_{2} - E_{1}$). Gamma-ray bursts (GRBs) and
TeV gamma-ray flares from active galactic nuclei (AGN) meet all of the
above criteria, and to date, have been used to place the strongest
limits on the energy scale of quantum gravity [H.E.S.S. Collaboration
  et al. 2011, Abdo et al. 2009].

\subsection{LIV with Gamma-ray Pulsars}
\cite{Karret1999} showed that gamma-ray pulsars were also excellent
tools for LIV searches and achieved a lower limit on the energy scale
of quantum gravity of $1.8\times10^{15}$~GeV for case of $n=1$
(Equation~2).  Further, doing LIV tests with gamma-ray pulsars has
several benefits over flaring transients such as GRBs or AGNs:
\begin{itemize}
\item Tests do not rely on random transient events and observational
luck. 
\item Limits improve with longer exposures.
\item Limits are based on highly significant measures of peak
  positions rather than on a handful of single photons.
\item Delay effects intrinsic to the source can be distinguished from
  LIV effects.
\item If a delay effect is measured, the delay separation will appear
  as a constant phase offset as the pulsar period lengthens, if it's a
  source effect.
\item If a delay effect is measured and is caused by LIV, this effect
  should stay constant in time as the pulsar slows and so should shift
  in phase by the amount $\Delta \Phi (t) = \Delta t/(P + t\dot{P})$.
\end{itemize}
Following the recent detection of the Crab pulsar by VERITAS, which
has extended the observed energy range of the pulsar by over an order
of magnitude, we are prompted to repeat the LIV test performed by
\cite{Karret1999}.

\subsection{Preliminary Results}
\begin{figure*}[t]
\subfigure[$\,\,\,$Crab pulsar main pulse]
{
\includegraphics[width=0.44\textwidth]{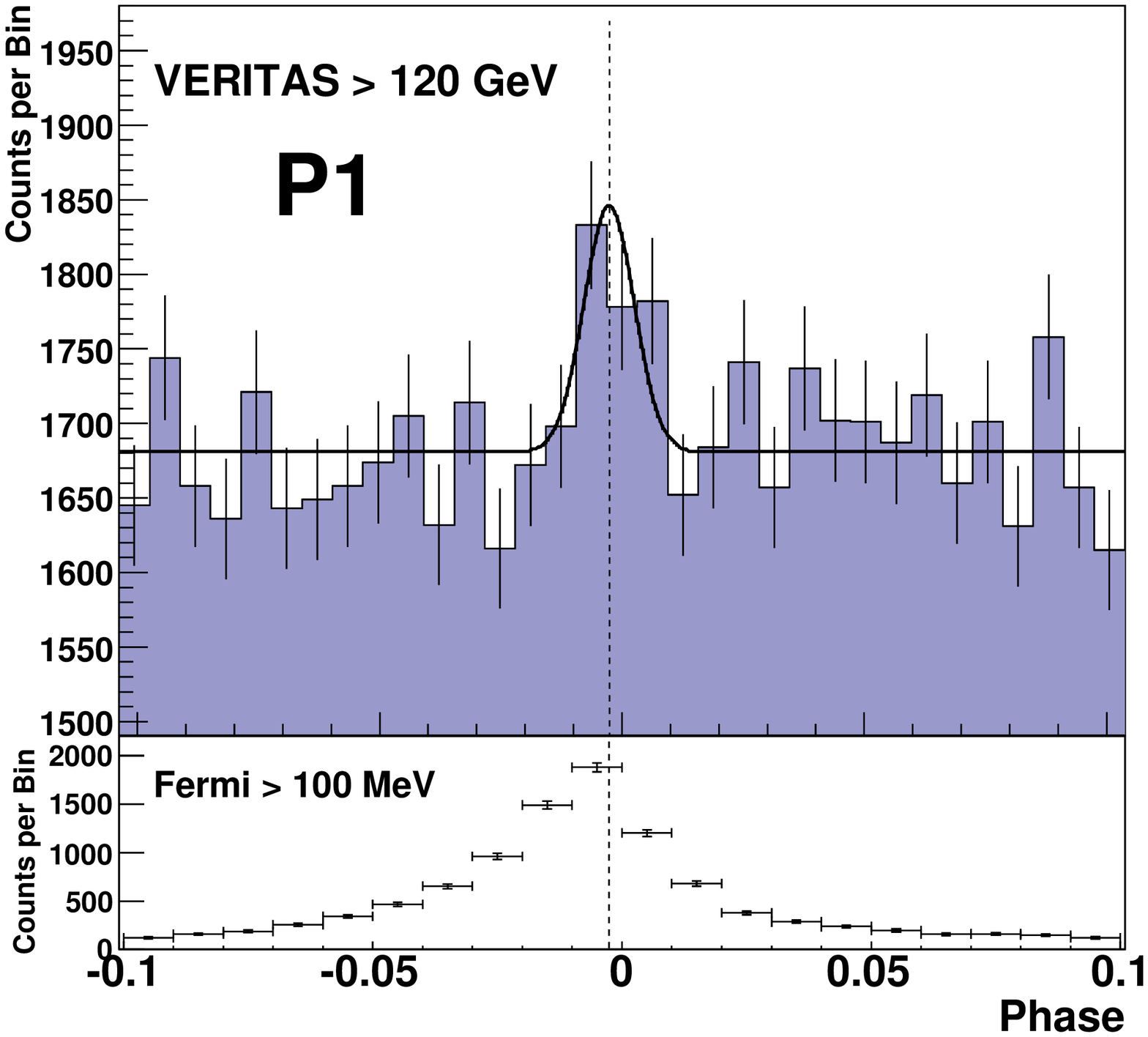}
}
\hspace{1.0cm}
\subfigure[$\,\,\,$Crab pulsar interpulse]
{
\includegraphics[width=0.44\textwidth]{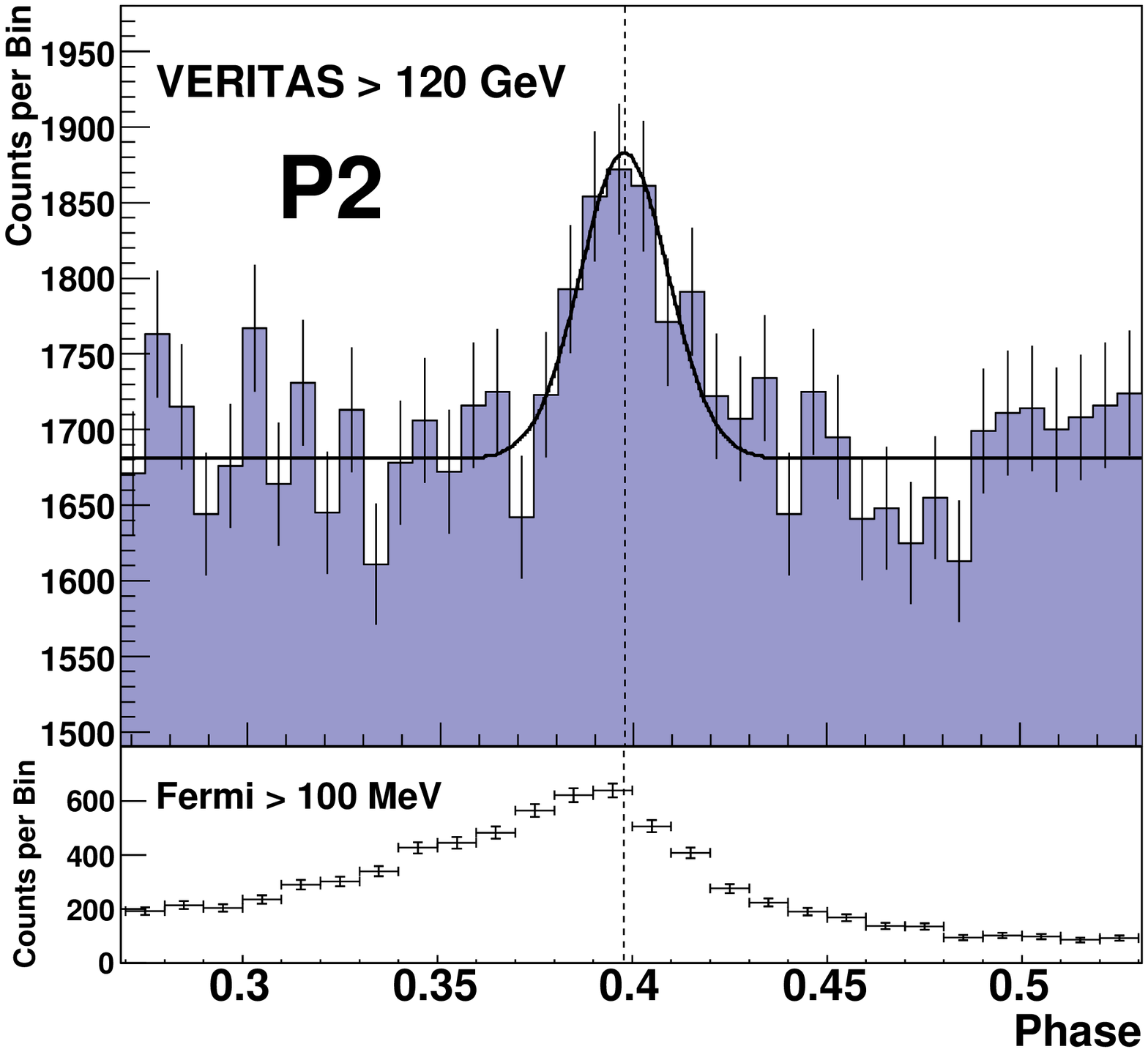}
}
\caption{A zoom in of the Crab pulsar gamma-ray emission peaks
  \citep{Aliu2011Sci}. The full VERITAS profiles are shown in
  Figure~1. The overlaid curve is from a maximum likelihood fit to the
  unbinned VERITAS data. The \emph{Fermi} points are taken from
  \cite{Abdo2010}.}
\end{figure*}
Using the VERITAS and \emph{Fermi} Crab pulse profiles shown in
Figure~4 we can place a limit on $E_{QG}$. The positions of the
emission peaks are consistent within measurement uncertainty. The
measurement of the VERITAS emission peak positions, which was
determined from an unbinned maximum likelihood fit to the VERITAS
phase data (solid line plotted in Figure~4), is the dominant source of
uncertainty.  This uncertainty is about $2\times10^{-3}$ phase units,
or, 60~$\mu s$. Using the 95\% confidence level upper limit on the
time separation between the peaks, 100~$\mu s$, the canonical
distance to the Crab pulsar of 2~kpc, and the conservative value of
120~GeV for the difference between the \emph{Fermi} and VERITAS energy
ranges, we arrive at $3\times10^{17}$~GeV and $7\times10^{9}$~GeV for
the lower limit on the values of $E_{QG}$ for the liner and quadratic
case, respectively.

\section{Conclusion}
VERITAS is engaged in a broad and varied pulsar observation
campaign. These proceedings present a summary of two research lines on
the Crab pulsar which have recently been, or are soon to be,
completed. These include the first limits on the level of enhanced VHE
emission correlated with GRPs in the Crab pulsar. Given the level of
uncertainty in theories of GRP emission, it is hard to draw firm
conclusions resulting from the lack of any observed correlation with
VHE emission. We are not aware of any theory with quantitative
predictions of correlated emission between GRPs and VHE
emission. Small and localised changes in the pair-creation rate, which
can explain the small (3\%) optical enhancements previously measured
by \cite{Shearer2003}, would yield VHE flux enhancements which are
below our sensitivity.

We also present some preliminary limits on the energy scale of quantum
gravity determined from VERITAS and \emph{Fermi} observation of the
Crab pulsar. These limits are only one order of magnitude lower than
the best limits ever achieved (with GRBs and AGNs) and are set to
improve with more data, more VHE pulsars, and improved analysis
methods.

\begin{acknowledgments}
This research is supported by grants from the U.S. Department of
Energy Office of Science, the U.S. National Science Foundation and the
Smithsonian Institution, by NSERC in Canada, by Science Foundation
Ireland (SFI 10/RFP/AST2748) and by STFC in the U.K.  We acknowledge
the excellent work of the technical support staff at the Fred Lawrence
Whipple Observatory and at the collaborating institutions in the
construction and operation of the instrument. The National Radio
Astronomy Observatory is a facility of the National Science Foundation
operated under cooperative agreement by Associated Universities, Inc.

This work was done in collaboration with Vlad Kondratiev who analysed
the GBT data. I am grateful to Nepomuk Otte and Ben Zitzer for their
help preparing these proceedings.

\end{acknowledgments}

\bigskip 

\end{document}